\documentclass[11pt,preprint]{aastex}

\usepackage{placeins}

\shorttitle{FU Ori rotation}
\shortauthors{Zhu \etal }

\begin{document}

\title{The Differential Rotation of FU Ori}

\author{Zhaohuan Zhu\altaffilmark{1},
Catherine Espaillat\altaffilmark{1},
Kenneth Hinkle\altaffilmark{2},
Jesus Hernandez\altaffilmark{1},\\
Lee Hartmann\altaffilmark{1}, and Nuria Calvet\altaffilmark{1}}
\altaffiltext{1}{Department of Astronomy, University of Michigan,
500 Church St., Ann Arbor, MI 48105} \altaffiltext{2}{National
Optical Astronomy Observatory, P.O. Box 26732, Tucson, AZ 85726}

\email{zhuzh@umich.edu, ccespa@umich.edu, hinkle@noao.edu, hernandj@umich.edu,\\
lhartm@umich.edu, ncalvet@umich.edu}

\newcommand\msun{\rm M_{\odot}}
\newcommand\lsun{\rm L_{\odot}}
\newcommand\msunyr{\rm M_{\odot}\,yr^{-1}}
\newcommand\be{\begin{equation}}
\newcommand\en{\end{equation}}
\newcommand\cm{\rm cm}
\newcommand\kms{\rm{\, km \, s^{-1}}}
\newcommand\K{\rm K}
\newcommand\etal{{\rm et al}.\ }
\newcommand\sd{\partial}
\newcommand\mdot{\rm \dot{M}}
\newcommand\rsun{\rm R_{\odot}}

\begin{abstract}
The emission of FU Orionis objects in outburst has been identified
as arising in rapidly accreting protoplanetary disks, based on a
number of observational properties. A fundamental test of the
accretion disk scenario is that the differentially rotating disk
spectrum should produce a variation of rotational velocity with the
wavelength of observation, as spectra taken at longer wavelengths
probe outer, more slowly rotating disk regions. Previous
observations of FU Ori have shown smaller rotation at near-infrared
($\sim 2.2 \mu$m) wavelengths than observed at optical ($\sim 0.6
\mu$m) wavelengths consistent with the assumption of Keplerian
rotation. Here we report a spectrum from the Phoenix instrument on
Gemini South which shows that differential (slower) rotation
continues to be observed out to $\sim 5 \mu$m. The observed spectrum
is well matched by the prediction of our accretion disk model
previously constructed to match the observed spectral energy
distribution and the differential rotation at wavelengths $\lesssim
2.2 \mu$m. This kinematic result allows us to confirm our previous
inference of a large outer radius ($\sim$1 AU) for the rapidly
accreting region of the FU Ori disk, which presents difficulties for
outburst models relying purely on thermal instability. While some
optical spectra have been interpreted to pose problems for the disk
interpretation of FU Ori, we show that the adjustment of the maximum
effective temperature of the disk model, proposed in a previous
paper, greatly reduces these difficulties.
\end{abstract}


\section{Introduction}

FU Orionis systems are a class of exceptionally luminous young
stellar objects found in star-forming regions (Hartmann \& Kenyon
1996). Originally identified from their very large increases in
optical brightness over timescales of years or less (Herbig 1977), a
larger group of heavily extincted probable members of the class have
been identified by their characteristic infrared spectra, which
strongly differ from those of typical T Tauri stars (Reipurth \&
Aspin 1997; Aspin \& Reipurth 2003; Reipurth \etal 2007). Additional
support for the identification of these heavily embedded objects
comes from recent high-resolution infrared spectroscopy, which
indicates that many of these sources are rapidly rotating and
exhibit double-peaked absorption line profiles as observed in FU Ori
(Greene, Aspin, \& Reipurth 2008).

The accretion disk model for FU Ori objects (Hartmann \& Kenyon 1996, and references therein)
rests fundamentally on the need to explain the peculiar spectral energy distributions (SEDs) of
these objects, which are much broader than that of a single temperature blackbody or star,
and which exhibit a continuously varying spectral type as a function of wavelength.
The disk model naturally accounts for these properties, as observations at longer wavelengths
probe increasingly cooler disk regions with later spectral types; our detailed model for
FU Ori matches the SED from optical wavelengths to the mid-infrared region (Zhu \etal 2007, 2008).
In addition, the disk model predicts that differential rotation should be observed, with
slower rotation seen at longer wavelengths, which arise from outer disk radii, and this
has been confirmed by comparing optical ($\sim 0.6 \mu$m) and
near-infrared ($\sim 2.2 \mu$m) spectral line profiles (Hartmann \& Kenyon 1987a,b;
Kenyon, Hartmann, \& Hewett 1988).

A desirable feature in a model or theory is an ability to make predictions that can
be tested observationally.  The disk model predicts that the observed rotational spectral
line broadening should be even smaller at $\lambda \sim 5 \mu$m (using the
fundamental CO vibrational transitions) than at $2.2 \mu$m (using the first overtone
CO vibrational lines).  In this paper we present a high-resolution spectrum of FU Ori in
the $5 \mu$m region which matches the predictions of the disk model.
We also show that some discrepancies seen in optical spectra of FU Ori in comparison
with simple disk models are alleviated by a decrease in the maximum disk temperature which
we proposed in Zhu \etal (2007).

\section{Observations}

A high-resolution spectrum of FU Ori at $4.9 \mu$m was obtained at
UT 01:00:29 on 2007 February 4 using the Phoenix spectrometer
(Hinkle \etal 1998, 2000, 2003) on the 8-m Gemini South telescope.
Observations were taken with a two-pixel slit
(0.17$^\prime$$^\prime$) for a resolution of
$\lambda/\delta\lambda=75,000$ over the wavelength range
$4.808-5.050 \mu$m.  We observed FU Ori at two positions along the
slit for eight 2 minute exposures.  We also observed the B2 III star
HR1790 for telluric line correction and took 10 flat-field and dark
images.

We reduced the data using IRAF.
\footnote{IRAF is distributed by the National Optical Astronomy Observatories,
which are operated by the Association of Universities for Research in Astronomy,
Inc., under cooperative agreement with the National Science Foundation.}
We averaged the flat-field and dark
images and subtracted the average dark image from the average flat-field image.  This averaged,
dark-subtracted flat-field image was then divided into the target spectra.
Images at different positions
of the slit were differenced to
remove the sky and dark backgrounds.  We then extracted the spectra using
the IRAF $apall$ routine and later combined and flattened the spectra using $splot$ in IRAF.
The spectrum of the hot star HR1790 was used to
divide out the telluric lines from the FU Ori spectrum.
Wavelength calibration was computed using telluric lines
from the Arcturus atlas of Hinkle \etal (1995).

We also obtained a high-resolution spectrum of FU Ori at UT 17:59:15
on 2008 November 14 using the Magellan Inamori Kyocera Echelle
(MIKE) spectrograph on the  6.5 m Magellan Clay telescope at Las
Campanas observatory (Bernstein \etal 2003).  MIKE is a double
echelle spectrograph which delivers full wavelength  coverage from
about $3350-5000$~\AA\ (blue side) and $4900-9500$~\AA\  (red side).
The data were obtained in subarcsecond  seeing with a 0.7\arcsec
slit, binning 2x2 and an exposure time of 120s. The resolutions were
$\sim$ 40,000 and $\sim $ 30,000 for the blue and red sides
respectively. The MIKE data were reduced using the MIKE Redux IDL
pipeline. \footnote{http://web.mit.edu/$\sim$burles/www/MIKE/}

\section{Results}

Figure 1 shows the reduced Phoenix spectrum of FU Ori.  Strong telluric features mean
that regions around wavenumber 2009.3 ${\rm cm^{-1}}$ are not usable, and small residuals from the telluric
correction can be seen at wavenumbers 2010.4, 2010.9, 2011.9, and 2013.5 ${\rm cm^{-1}}$.
The spike around wavenumber 2012.5 ${\rm cm^{-1}}$ is due to a bad pixel.
Outside of these regions, an absorption spectrum is clearly
present, with relatively broad features and substantial blending.  Most of the lines are
due to the P-branch of the fundamental rotational-vibrational transitions of CO.

To interpret the results, we calculated a synthetic disk spectrum
using the methods described by Zhu \etal (2007).  The model
parameters were those used by Zhu \etal (2007) to fit the SED of FU
Ori and to match the observed rotational broadening observed at
optical and near-infrared wavelengths:
central star mass $\sim$ 0.3 M$_{\odot}$,
mass accretion rate of the inner high $\dot{M}$ disk
$\sim$2.4$\times$10$^{-4}\msunyr$, disk inner radius $\sim$5
$R_{\odot}$, outer radius of the inner high $\dot{M}$ disk $\sim$ 1
AU, and the disk inclination angle $\sim$55$^{o}$.
(See Figure 8 of Zhu \etal for the fit to the $2.2 \mu$m CO lines).
We emphasize that
we have {\em not} changed or adjusted any parameters from the Zhu
\etal (2007) FU Ori model; these are predicted spectra.

The lower dotted curve in Figure 1 shows the synthetic disk spectrum
observed pole-on, so that individual spectral lines can be seen
without the blending that occurs due to the large rotational
broadening. Comparison of the nonrotating spectrum with the
observations shows that FU Ori has significantly larger line widths,
consistent with rapid rotation, and unlike profiles of M giants and
supergiants. The upper dotted curve shows the synthesized spectrum
using the inclination and central mass used to obtain a match to the
$2.2 \mu$m CO line widths. The agreement between synthetic and real
spectra is quite good, except near $2010.7 {\rm cm^{-1}}$ where we
are missing the CO 5-4 P7 line in the model. This could be due to
adopting too small an oscillator strength \footnote{This work
employed the line lists from Kurucz CD ROM-15}. The half-width at
half-depth (HWHD) of the lines in this spectral region (5 $\mu$m) is
$\sim 22 \kms$, considerably smaller than the line widths measured
at $2.2 \mu$m (HWHD $\sim$ 36 km s$^{-1}$) (Hartmann \& Kenyon
1987a; Hartmann, Hinkle, \& Calvet 2004; Zhu \etal 2007). This HWHD
at 5 $\mu$m is close to the Keplerian velocity at 0.5 AU around 0.3
M$_{\odot}$ central star. All the strong lines have been identified
in the pole-on model spectrum. Though some water lines are present
in this spectrum, they are washed out or blended in the broadened
spectrum (upper dotted curve). Only the strong CO fundamental lines
can be identified with certainty.

It is also worth noting that some of the unblended line profiles
exhibit evidence for double-peaked shapes predicted by simple disk
models (e.g., Hartmann \& Kenyon 1987a,b). In addition, some of the
blends show sharp features (eg. 2011.2 cm$^{-1}$ line) which are the
result of overlapping double-peaked lines.  These features naturally
arise in a disk model but would not be seen in rotating star models
(unless large polar spots are invoked; see below).

In Figure 2 we display a segment of the MIKE spectrum of FU Ori in
the wavelength range $7030 - 7100$~\AA\ for comparison with the
synthetic Keplerian disk spectrum.  We again find good agreement
between model and observation, demonstrating that there has been no
change in the estimated optical rotational velocity of the object
between the observations in Zhu \etal 2007, which we used to set
the disk parameters, and this paper.
The HWHD of the optical lines in this wavelength range
is $\sim$65$\pm$5 km s$^{-1}$, consistent with HWHD
$\sim$ 62$\pm$5 km s$^{-1}$ measured by Petrov \& Herbig (2008; PH08).

Thus, compared with HWHD of 2 micron CO first-overtone lines $\sim$
36$\pm$3 km s$^{-1}$ and HWHD of 5 micron CO fundamental lines
$\sim$ 22$\pm$2 km s$^{-1}$, the differential rotation in FU Ori
observed over nearly an order of magnitude in wavelength is
consistent with Keplerian rotation.  The slow rotation observed at
$5 \mu$m implies spectral formation at radii out to $\sim 0.5$~AU,
in agreement with our SED modeling; this supports our conclusion in
Zhu \etal (2007, 2008) that the extent of the high-accretion rate
disk is larger than can be explained by pure thermal instability
models for outbursts \citep{1994ApJ...427..987B}.

\section{Discussion}

The consistency of the variation of rotational velocities as
observed over to $\lambda \sim 0.7 - 5 \mu$m with Keplerian rotation
seemingly provides strong evidence for the accretion disk
interpretation of FU Ori.  However, PH08
recently argued that while the infrared spectrum is that of an
accretion disk, the optical spectrum is produced by a
rapidly-rotating star with a dark polar spot.  The PH08 argument
against a pure disk model for a central star rests on three main
points: there is no evidence for a variation of absorption line
width as a function of excitation potential over the wavelength
range $\lambda \sim 0.52 - 0.86 \mu$m; there is no evidence for a
variation of rotational velocity with wavelength over that
wavelength range; and the observed line profiles are more ``boxy''
or flat-bottomed than the double-peaked profiles of the disk model.

It is important to recognize that the above-listed effects expected
for a disk spectrum require not only differential rotation but a
temperature gradient as well. A Keplerian disk exhibiting a constant
effective temperature would not show any effect of rotational
velocity with either excitation or wavelength; and the double-peaked
behavior of line profiles only occurs because the outer, slowly
rotating regions do not fill in the profile at line center, as these
regions are too cool to emit significantly at the wavelength of
observation. While the standard steady disk temperature distribution
$T_{eff}^4 \propto [1 - (R_i/R)^{1/2}] R^{-3}$ is a power law at
large radii, it is relatively flat at distances within about twice
the inner radius $R_i$. Therefore, observations at long wavelengths
which probe the outer disk where the temperature falls rapidly with
radius will exhibit stronger rotational velocity variations and more
double-peaked line profiles than observations at short wavelengths
probing the inner, more nearly isothermal disk.

In Zhu \etal (2007) we were forced to use a
maximum disk temperature $T_{max} = 6420$~K to match the SED of FU
Ori, which is lower than the 7200 K maximum temperature adopted in
the model used by PH08 (which was based on the earlier model
by Kenyon, Hartmann, \& Hewett 1988).  Lowering the maximum temperature
has the effect of making the flatter
part of the accretion disk temperature distribution more dominant at
optical wavelengths. As shown in Table 1 and Figure 3, this model
predicts essentially no variation of line width with lower level excitation
potential and a very slight dependence on wavelength in the optical region.
(Note that PH08 predict a
much larger effect of line width on excitation potential than
Welty \etal (1992) for what should be essentially the same disk
model; the reason for the large discrepancy is unclear.)
In any case, measurement of rotation is best done through cross-correlation
using suitable templates, as many of the lines used by PH08 are blends and
introduce very large scatter into the model predictions
(see Table 1 and Figure 3).

PH08 also noted that their disk model predicts very strong TiO
absorption bands at $\sim 7054$ and $7087$~\AA\ which are not
observed.  However, as shown in Figure 2, our lower-temperature disk
model does not predict strong TiO absorption bands in this region.
In addition there is evidence for $7087$~\AA\ bandhead absorption in
our MIKE spectra, at the level predicted by our disk model.  Once
again this difference in the predicted disk model spectra arises
simply by reducing the maximum disk temperature, which increases the
importance of hot inner disk relative to the outer cool disk at the
wavelength of observation.

It has long been known that many optical line profiles in FU Ori are less double-peaked
than predicted by simple quiescent disk models (Hartmann \& Kenyon 1985).
There are, however, alternative possibilities to explain the profiles which do not
demand abandonment of the accretion disk hypothesis.
If, as currently thought, ionized disks accrete through the action of the magnetorotational
instability (MRI; Balbus \& Hawley 1998), such disks must be turbulent.
The disk models with resolved
vertical structure computed by Miller \& Stone (2000) predict that turbulence
driven by the MRI in the central layers of the disk produces waves which propagate outward
and shock in the upper layers.   It would be surprising if the MRI did not
produce significant turbulent motions in the upper atmospheric layers of the disk,
which would tend to wash out the double profile structure.
Hartmann, Hinkle, \& Calvet (2004) found that
that some mildly supersonic turbulence was needed to explain the $^{12}$CO first-overtone
lines of FU Ori.

It should be noted that the standard steady disk structure may not
be completely applicable in the innermost disk. Standard thin-disk
models predict that accretion onto a slowly rotating star should
give rise to boundary-layer emission with roughly half the system
luminosity; this is not observed in FU Ori (Kenyon \etal 1989).
Popham \etal (1996) considered disk models which suppress boundary
layer radiation; such models exhibit less doubled line profiles in
inner disk regions, largely because the angular velocity of the disk
departs from Keplerian values near the inner disk boundary.

To explain the optical spectrum of FU Ori with a central star, the
star would be required to have essentially the same total system
luminosity $L \sim 230 \lsun$, and would need to have a radius
roughly twice the inner radius of the disk model, $R \sim 10 \rsun$
(Zhu \etal 2007).  Assuming Keplerian rotation for the infrared
disk, Zhu \etal estimated a central mass $M \sim 0.3 \msun$. Such a
star cannot be an isolated product of stellar evolution, as it has
an implausibly short Kelvin-Helmholtz contraction time $\sim G M^2
R^{-1} L^{-1} \sim 1200$~yr.  The energy to power the star would
have to come from disk accretion, which would also potentially
explain the outburst (Larson 1983). However, as the ratio of
optical-to-infrared rotational velocities is consistent with a
Keplerian profile, this implies that any central star would have to
be rotating nearly at breakup; this means that the assumption of
solid-body rotation in the PH08 model is unlikely to be correct. It
is also unclear whether the accretion of a large amount of hot disk
material would add enough angular momentum to spin up the outer
layers of the star to breakup velocity as it expanded the outer
atmosphere.

In summary, the accretion disk model for FU Ori provides a coherent
explanation of the observed spectral energy distribution and
differential rotation over more than a decade in wavelength. The
slow rotation observed at $5 \mu$m supports our previous result that
the high mass accretion rate disk could extend to $0.5-1$~AU, which
is significantly larger than that predicted by the pure thermal
instability theory \citep{1994ApJ...427..987B}. On the other hand,
the theory incorporating both gravitational and magnetorotational
\citep{1999ASPC..160..122G,2005AAS...207.7417B,2009} successfully
predicts AU scale high mass accretion rate inner disk during
outbursts \citep{2009}. With the advent of more powerful computers
and sophisticated magnetohydrodynamic codes, and the assumption of
MRI-driven accretion, it should be possible to explore the
possibility that atmospheric turbulence and/or nonstandard inner
thin disk structure can explain details of the optical line
profiles.

This work is supported in part by NASA grant NNX08AI39G and is based
in part on observations obtained at the Gemini Observatory, which is
operated by the Association of Universities for Research in
Astronomy, Inc., under a cooperative agreement with the NSF on
behalf of the Gemini partnership: the National Science Foundation
(United States),the Science and Technology Facilities Council
(United Kingdom), the National Research Council (Canada), CONICYT
(Chile), the Australian Research Council (Australia), Ministrio da
Cincia e Tecnologia (Brazil) and Ministerio de Ciencia, Tecnologa e
Innovacin Productiva  (Argentina). The observations were obtained
with the Phoenix infrared spectrograph, which was developed and is
operated by the National Optical Astronomy Observatory.  The
Gemini/Phoenix spectra were obtained as part of program
GS-2007A-C-4.

\begin{table}
\begin{center}
\caption{HWHD of lines selected by PH08 as measured from Our
synthetic disk spectrum\label{linewidth}}
\begin{tabular}{ccccc|ccccc}
\tableline\tableline
$\lambda$& Ion  & EP & HWHD & Grade \tablenotemark{a}& $\lambda$& Ion & EP & HWHD & Grade \\
(\AA)&  & (eV) & (km s$^{-1}$) &  & (\AA) & &
(eV) & (km s$^{-1}$) &\\
\tableline
5383.37 & Fe I & 4.31 & -&0 & 6726.66 & Fe I & 4.61 &66.2 &3  \\
5717.83 & Fe I & 4.28 & -&1 & 6767.79 & Ni I & 1.83 &65.4 &3  \\
5772.15 & Si I & 5.08 & -&0 & 6810.26 & Fe I & 4.61 &- &1  \\
5775.08 & Fe I & 4.22 & -&1 & 6814.94 & Co I & 1.96 &- &0  \\
5862.35 & Fe I & 4.55 & -&0 & 6828.59 & Fe I & 4.64 &78 &3  \\
5899.29 & Ti I & 1.05 & 61.3&2 & 7090.38 & Fe I & 4.23 & 66.8 & 4 \\
5922.11 & Ti I & 1.05 & 61.8&3 & 7122.19 & Ni I & 3.54 & 66.1 & 4  \\
5934.66 & Fe I & 3.93 & 78.4&3 & 7344.70 & Ti I & 1.46 & 76.8& 2  \\
5965.83 & Ti I & 1.88 & 65.9&3 & 7393.60 & Ni I & 3.61 & 68& 3  \\
5987.07 & Fe I & 4.80 & 71.7&3 & 7445.75 & Fe I & 4.26 & -&0  \\
6016.66 & Fe I & 3.55 &- &1 & 7511.02 & Fe I & 4.18 & 61.7 & 4  \\
6024.06 & Fe I & 4.55 &- &1 & 7525.11 & Ni I & 3.63 & -&1  \\
6027.05 & Fe I & 4.09 &- &0 & 7555.60 & Ni I & 3.85 & 68.3&3  \\
6056.01 & Fe I & 4.73 &- &1 & 7568.89 & Fe I & 4.28 & 69.4&4  \\
6108.11 & Ni I & 1.68 &- &1 & 7574.04 & Ni I & 3.83 & -&1  \\
6180.20 & Fe I & 2.73 &70 &2 & 7586.01 & Fe I & 4.31 & 65.8&5  \\
6270.23 & Fe I & 2.86 &72.2 &3 & 7727.61 & Ni I & 3.68 & 65.4&4  \\
6355.03 & Fe I & 2.85 &64.4 &3 & 7780.55 & Fe I & 4.47 & 65.4&4  \\
6358.70 & Fe I & 0.86 &66.05 &4 & 7788.94 & Ni I & 1.95 & 64.13&4  \\
6380.74 & Fe I & 4.19 &72.2 &3 & 7855.44 & Fe I & 5.06 & 62.1&4  \\
6411.65 & Fe I & 3.65 &67.8 &2 & 7912.87 & Fe I & 0.86 & -&0  \\
6439.08 & Ca I & 2.53 &74.5 &3 & 7937.13 & Fe I & 4.33 & 63.5&4  \\
6471.66 & Ca I & 2.53 & 72.8 &3 & 8075.15 & Fe I & 0.92 & -&0  \\
6475.62 & Fe I & 2.56 &66.3 &5 & 8080.55 & Fe I & 3.30 & -&0  \\
6569.22 & Fe I & 4.73 &67.8 &5 & 8085.18 & Fe I & 4.45 &64.7 &3  \\
6581.21 & Fe I & 1.49 &64.7 &4 & 8426.51 & Ti I & 0.83 &- &0  \\
6586.31 & Ni I & 1.95 & -&1 & 8611.80 & Fe I & 2.85 &65.0 &4  \\
6707.89 & Li I & 0.00 & -&0 & 8621.60 & Fe I & 2.95 &63.5 &3  \\
6717.68 & Ca I & 2.71 & 69.0&4 & 8648.47 & Si I & 6.21 & 73.9 & 3 \\
6721.85 & Si I & 5.86 & 65.2&2 &   \\

\tableline
\end{tabular}
\tablenotetext{a}{HWHD means "half-width at half-depth" as
illustrated and defined in Figure 3 of PH08. Grade indicates the amount
of line blending due to rapid rotation:
grades 0 and 1 signify badly blended lines,
grade 2 identifies less-blended lines that are difficult to use,
and grade 3 means that the line is moderately blended so that the true HWHD may
be slightly smaller than our measured value to within 10 km s$^{-1}$.
Grade 4 means the line is slightly blended with an errorbar of only 3 km s$^{-1}$,
while grade 5 means that the line is blend-free. }
\end{center}
\end{table}

\begin{figure}
\includegraphics[width=0.9\textwidth]{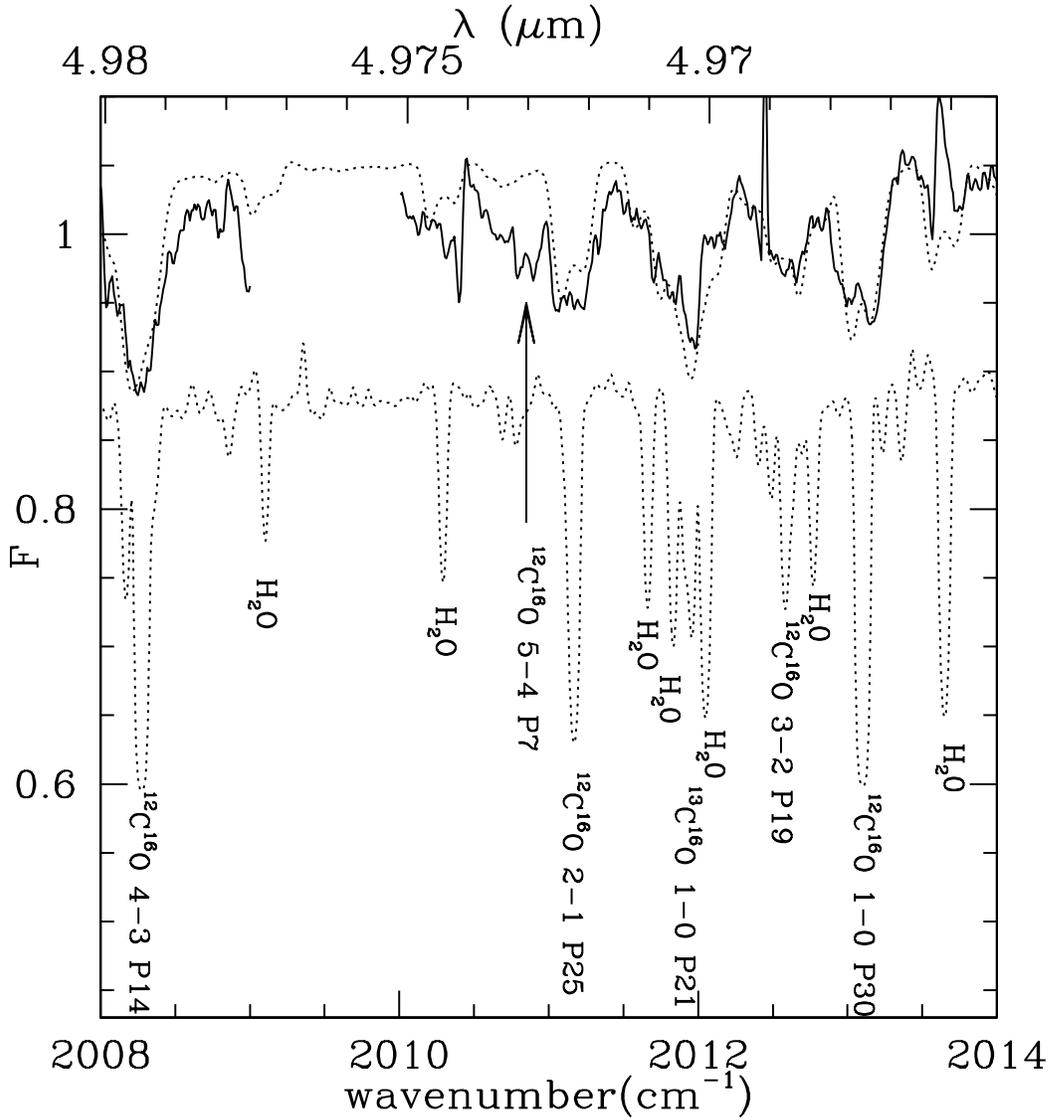}

\label{fig:f1} \caption{Observed 4.9~$\mu$m Phoenix spectrum of FU
Ori (solid curve) compared with our synthetic disk model spectrum
(upper dotted curve). While the wavenumber is shown at the bottom,
the wavelength in microns is shown at the top of the figure. The
lower dotted curve shows the model spectrum observed pole-on, which
removes the rotational broadening, illustrating the need for rapid
rotation to match the observations (see the text). The strong lines
are identified in the lower model spectrum. The arrow points to the
strong CO line which is missing in our model.}
\end{figure}

\begin{figure}
\includegraphics[width=0.9\textwidth]{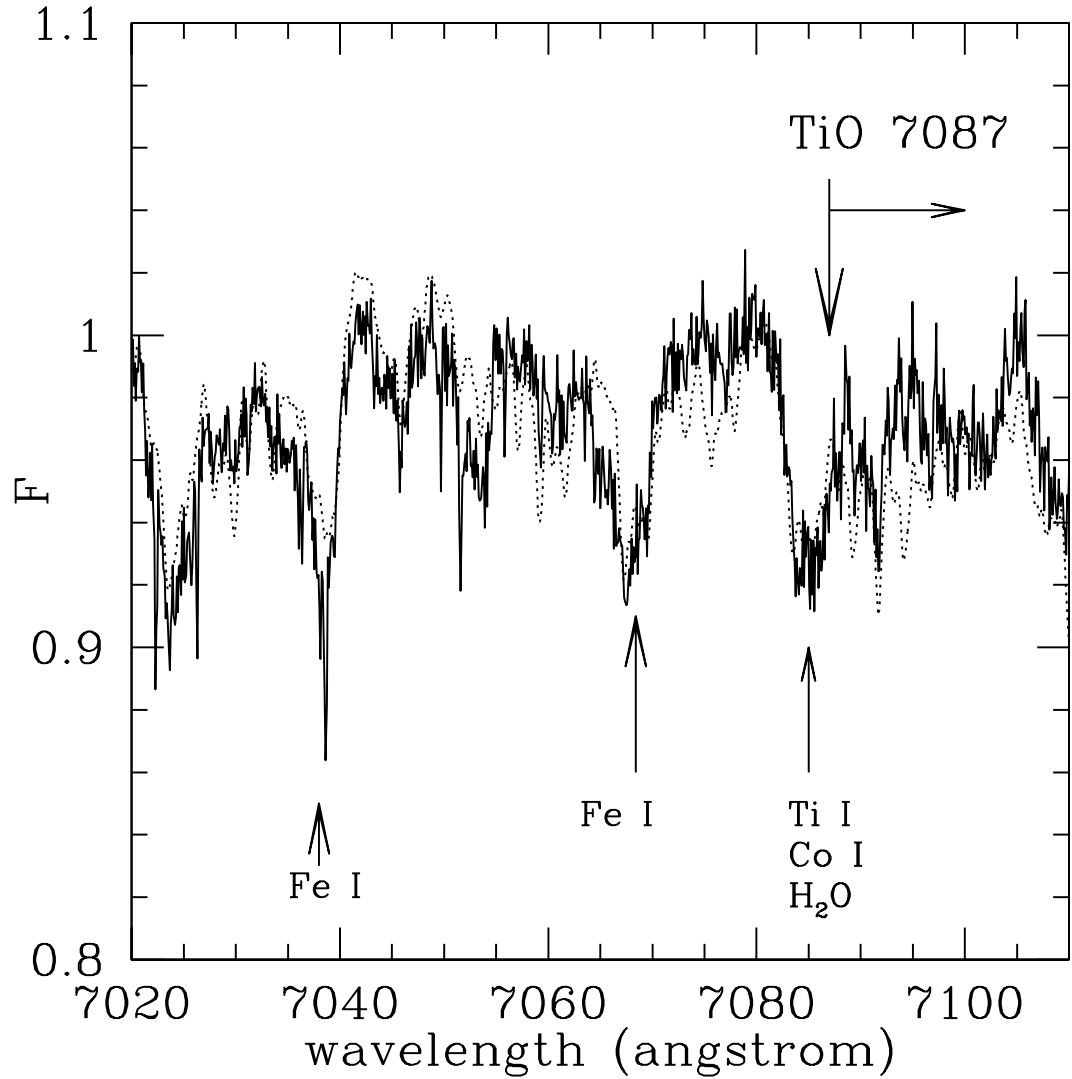}
\label{fig:f2} \caption{Comparison of the observed MIKE spectrum of
FU Ori near 7080 \AA\ (solid curve) with the synthetic disk spectrum
(dotted curve). Compared with the continuum flux $\sim$ 7080 \AA,
there is some evidence for weak TiO 7087 \AA\ band absorption in
both the observed and the model spectrum (see text).}
\end{figure}

\begin{figure}
\includegraphics[width=0.42\textwidth]{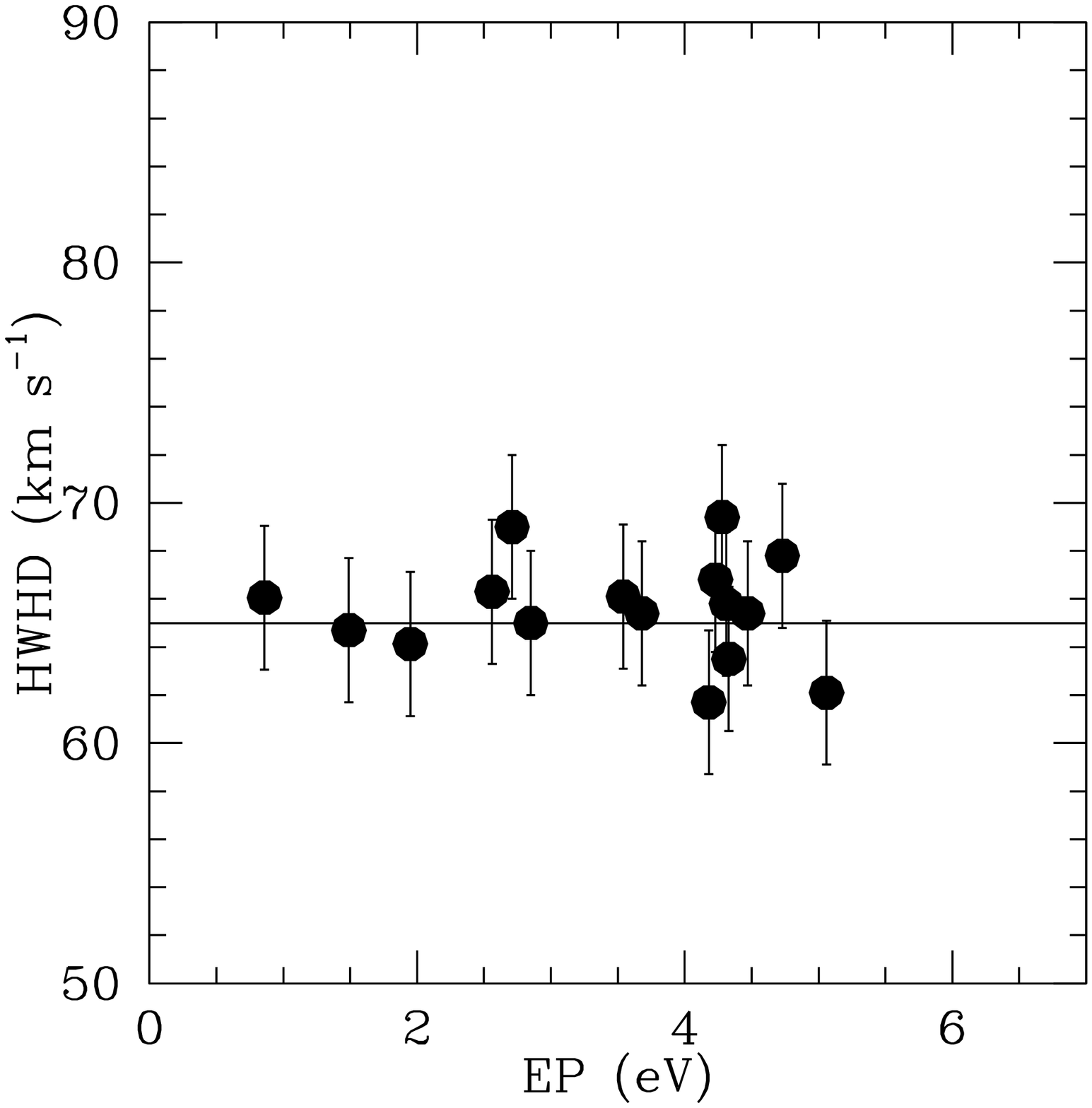} \hfil
\includegraphics[width=0.42\textwidth]{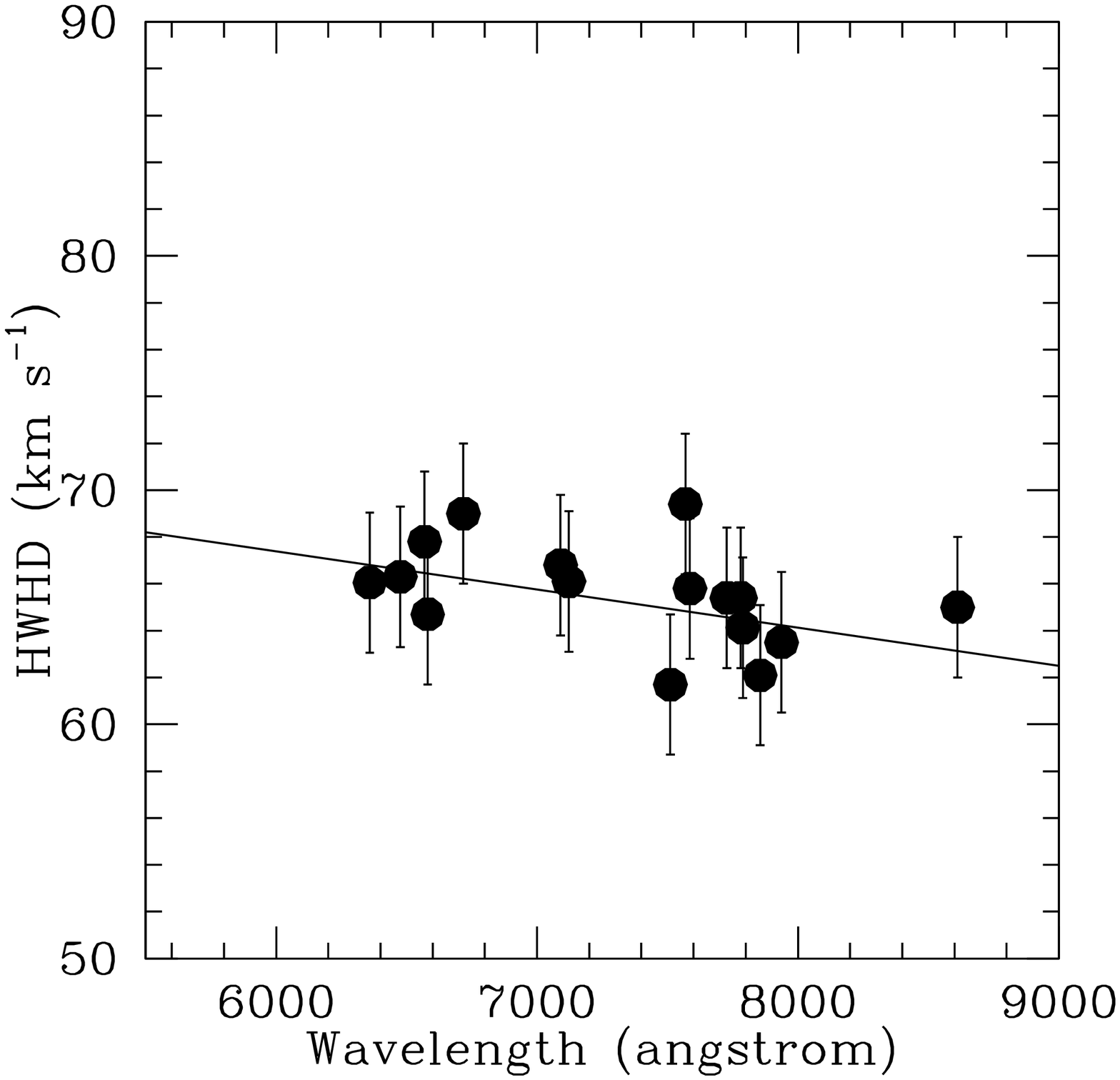} \\
 \caption{Dependence of the model line half-width at half-depth (HWHD)
on excitation potential (EP) (left) and the wavelength (right),
taken from Table 1. We show only the grade 4 and 5 line widths for
clarity (see Table 1 for their definitions). }
\end{figure}

\end{document}